**Oxygen stoichiometry and instability in aluminium oxide tunnel barrier layers**

E. Tan, P.G. Mather, A.C. Perrella, J.C. Read and R.A. Buhrman

Cornell University, Ithaca NY 14853-2501

ABSTRACT

We present X-ray photoelectron spectroscopy data which show that the chemisorbed oxygen previously observed to be on the surface of thin $AlO_x$ layers formed by room temperature thermal oxidation is bound by oxygen vacancies in the oxide. Increasing the electric field across the oxide, either by over-coating with a metallic electrode, or by electron bombardment, drives this surface chemisorbed oxygen into the vacancy sites. Due to the low bonding energies of these oxygen sites, subsequent oxygen exposures draw these $O^-$ ions back to the surface, reforming chemisorbed $O_2^-$. $Al/AlO_x/Al$ tunnel junctions incorporating electron bombarded $AlO_x$ barriers show a significant reduction in the low frequency junction resistance noise level at 4.2 K.

The use of aluminum oxide (AlO$_x$) layers of nanometer thickness formed by ~300 K oxidation of aluminum thin films has long been the most successful approach to the fabrication of high-performance metal-insulator-metal tunnel junctions, originally for low temperature superconducting Josephson junctions[1] (JJs), and more recently for magnetic tunnel junctions[2] (MTJs). The goals of reliably forming still thinner barriers to achieve higher critical current densities $J_c$ (for JJs) and lower specific resistances (for MTJs), of producing tunnel junctions with lower levels of 1/$f$ noise for sensor and quantum computing applications, and of forming higher performance gate insulators for molecular electronics studies[3] have continued to focus attention on the objective of obtaining a better understanding and improved control of the oxide barrier layer. Recently a scanning tunneling and ballistic electron emission microscopy (STM/BEEM) study[4] of the thermal oxidation of aluminum revealed that after oxidation, a layer of oxygen remains on the surface of the oxide indefinitely in ultra-high vacuum in the form of nanoscale clusters of chemisorbed $O_2^-$, raising questions regarding the cause of this chemisorbed oxygen and its effect on the resultant tunnel barrier when over-coated by the counter electrode.

Here we report on an X-ray photoemission spectroscopy (XPS) study of the thermal oxidation of Al, which extends previous XPS studies[5-6] and shows that this chemisorbed oxygen is associated with and bound to the surface by O vacancies in the oxide. Low energy electron bombardment can drive the chemisorbed oxygen into the oxide layer and substantially stabilize it there but subsequent exposure to oxygen ambient pulls some of this oxygen back to the surface, demonstrating the low bonding energies of a portion of the oxygen sites in the oxide. Depending on the relative work function $\Phi$ of the metal used, much of this chemisorbed oxygen is also driven into the oxide layer upon over-coating by a metallic counter electrode. These results yield new insights into the structure and electronic properties of thermally formed AlO$_x$ layers and suggest new pathways whereby thinner, more stable, and potentially lower 1/$f$ noise and higher electrical resistance tunnel barriers and gate insulators might be produced.

The polycrystalline Al thin film samples in this investigation were produced by thermal evaporation onto an oxidized Si substrate in an ultra-high vacuum environment (~$2 \times 10^{-9}$ torr). The surfaces were initially oxidized through a controlled time exposure to a reduced atmosphere of 99.998% pure, dry $O_2$ and then transferred via a "vacuum suitcase" (~$2 \times 10^{-9}$ torr) to an XPS system for analysis, further controlled oxidation and processing. The oxygen exposures employed (between $10^{-1}$ torr·s and $10^6$ torr·s) cover the range that produces Nb-Al-AlO$_x$-Nb JJs whose critical current density $J_c$ decreases, with increasing exposure[1], from $10^5$ A/cm$^2$ to $10^2$ A/cm$^2$. The oxygen pressures employed were 50 – 100 mtorr for the low doses ($\leq$ 10 torr·s) and 0.5 – 1 torr for the higher doses. As a thick Al layer was deposited, none of the oxygen exposures completely oxidized the Al, as is the case with the standard Nb JJ fabrication process. The XPS spectra were acquired using a Surface Science Laboratories SSX-100 Spectrometer[7], with the hemispherical pass energy kept at 25 eV.

Fig. 1a(i) shows the Al 2p photoemission spectrum from an Al film exposed to 5 mtorr·s of $O_2$. The effective thickness $t_{eff}$ of the oxide layer was determined from the relative intensity of the metal and oxide signals, using the standard value[8] for the mean free path (*mfp*) of an Al 2p photoelectron in α-Al$_2$O$_3$ and a reduced aluminium oxide density[9] of 3.4 g/cm$^3$. Fig. 1b(i) shows the corresponding O 1s spectrum, which is broad and asymmetric. A flood gun electron bombardment (FGEB) of the oxidized surface provides information to properly fit this spectrum. Fig. 1a(ii) and 1b(ii) shows the changes that a 1 hour FGEB, at a bias voltage of 20 V and an electron current density of ~10 $\mu$A/cm$^2$, effects on the O 1s XPS signal. The high binding energy portion of the signal is almost completely removed, while there is a substantial increase in the intensity of the low binding energy component. This facilitates the identification of the oxygen signal as consisting of a low binding energy component, centered at 532.6±0.1 eV, attributable to the oxide layer, and a component at 533.9±0.1 eV, attributable to chemisorbed oxygen.

With these identifications, we determined both $t_{eff}$ and the stoichiometry ($O_{oxide}/Al_{oxide}$ ratio) of a thermal oxide layer as an Al sample received oxygen doses that cumulatively cover the range from 0.5 to $10^6$ torr·s. In Fig. 2 we show $t_{eff}$ and the $O_{oxide}/Al_{oxide}$ ratio of one such layer as a function of oxygen dose $D_{ox}$, along with the $O_{chem}$ intensity, normalized to its value at 0.5 torr·s. Over this exposure range, $t_{eff} \propto \log D_{ox}$, as in the classic model of the initial stage of oxidation[10], before the rate of growth began to slow above $10^5$ torr·s. The nominal $O_{oxide}/Al_{oxide}$ ratio during this growth process remained in the range of 1.0 to 1.1. For $10^4 - 10^5$ torr·s exposures, this ratio dropped significantly to ~0.8, far below the stoichiometric value of 1.5. We note here that the $O_{oxide}/Al_{oxide}$ ratio was determined using the standard values[8] *mfp*s of the O 1s and Al 2p photoelectrons in $\alpha$-$Al_2O_3$, and the O/Al intensity ratio from an $\alpha$-$Al_2O_3$ reference sample. Since the *mfp* may be different in $AlO_x$, this stoichiometry measurement must be considered a relative measurement, not an absolute determination. The amount of chemisorbed oxygen $O_{chem}$ remained constant for initial and mid-range exposures ($10^4$ torr·s and below), but then approximately doubled during the final high dose exposures (above $10^4$ torr·s). Very similar results were obtained with a number of samples.

From the XPS spectra we determined that a FGEB such as that used to obtain the changes shown in Figs. 1a & 1b caused an increase in the effective thickness of the oxide from 0.89 nm to 1.18 nm, as well as an increase in $O_{oxide}/Al_{oxide}$ from 0.9±0.1 to 1.2±0.1.. If the oxide was briefly exposed to oxygen ambient, it reverted back to a more oxygen deficient form, with the reappearance of the chemisorbed peak, but whose intensity is lower than before. This is shown in Fig. 1b(iii).

In Fig. 3 the resultant oxide thickness and the $O_{oxide}/Al_{oxide}$ ratio are plotted for a series of 1 hour, 20 V flood gun bombardments and oxygen re-exposures (the first at 300 torr·s and the second at $10^4$ torr·s). Each bombardment was nearly completely effective in eliminating the chemisorbed oxygen component and in raising the oxide stoichiometry to $O_{oxide}/Al_{oxide}$ 1.2, but each oxygen exposure resulted in some of the previously incorporated oxygen moving back out of the oxide, lowering the oxide stoichiometry but not necessarily $t_{eff}$. Five different samples were subjected to multiple FGEBs in this manner with the resultant oxide stoichiometry being quite reproducible, $O_{oxide}/Al_{oxide} = 1.2 \pm 0.03$, but the oxygen loss upon exposure to oxygen ambient was more variable.

In the Mott-Cabrera model[10] of low temperature oxidation, the electric field across the oxide that results from chemisorbed $O^-$ ions drives the diffusion of Al cations from the metal through the oxide. The decreasing field strength with increasing oxide thickness results in the logarithmic growth rate. This basic model does not consider effects arising from defects and disorder in the oxide layer. Due to the strong bond length mismatch[11] between Al ions in the metal (0.265 nm for the Al(111) surface) and in the oxide (~0.32 nm) a stoichiometric $Al_2O_3$ layer, particularly as formed at ~300K, would be highly strained. X-ray and neutron scattering[11] and electron energy loss spectroscopy studies[12] have shown that the aluminium in $AlO_x$ is principally tetrahedrally coordinated, as would be expected from random filling of the interstitial sites of a randomly closed packed oxygen sub-lattice, rather than octahedrally coordinated as in crystalline $\alpha\text{-}Al_2O_3$, and considerably less dense[9] than $\alpha\text{-}Al_2O_3$.

The response to the electron bombardments and subsequent brief oxygen exposures indicates that the bonding energy of some of the oxygen sites within the oxide is comparably low. Thus we propose as the mechanism for the reduction in $O_{oxide}/Al_{oxide}$ of FGEB oxides upon oxygen ambient exposure that when a chemisorbed $O^-$ ion, generated by the dissociation of incident $O_2$ molecules, is located above such a site, it is energetically favorable for a weakly bonded oxygen ion within the oxide to be displaced to the surface to form $O_2^-$, which is then held onto the surface by the resultant positively-charged oxygen vacancy in the oxide. In this model, sufficiently strong, negative surface charging from electron bombardment causes $O_2^-$ to dissociate and drives the resultant $O^-$ ions back into these vacancies.

The oxygen vacancies in the as-grown $AlO_x$ layer and the associated chemisorbed $O_2^-$ bound to the surface by positively charged vacancies provide a mechanism for the substantial slowing down of the oxide growth below the Mott-Cabrera logarithmic rate. The approximate doubling in the chemisorbed signal and the decrease in oxide stoichiometry, once $t_{eff}$ reaches 13 – 1.5 nm, suggests that in this thickness range, the strain in the oxide becomes such that there is a substantial increase in the density of low binding energy oxygen sites within the oxide. The result is a large increase in the chemisorbed $O_2^-$ coverage, which then passivates most, if not all, of the oxide surface. The fact that chemisorbed oxygen is also found in XPS measurements of other thin and adherent oxides formed at ambient temperatures, such as cobalt oxide, suggests that this passivation mechanism may be fairly general. Indeed, under FGEB, thin oxidized cobalt samples behave similarly to the $AlO_x$ samples, with the chemisorbed O decreasing in amplitude, and the thickness and oxygen concentration of the cobalt oxide increasing. However, for Co-$CoO_x$ a 20 V flood-gun bias was insufficient to fully remove all the chemisorbed oxygen.

We examined what happens to the oxide system when it is over-coated by preparing samples consisting of a thick Al layer exposed to 1 torr·s $O_2$, but with different metal over-layers deposited onto the $AlO_x$ layer. The over-layer was thin enough (1 – 1.5 nm) that the resultant effective thickness of the oxide could be determined from the Al 2p spectra. BEEM experiments[13] have shown that this over-layer thickness provided a complete metallic coverage of the oxide. In Fig. 4 we plot the effective thickness of the oxide layer versus the work function[14] of the top electrode. For the case of an yttrium over-layer there was no increase in the oxide thickness; for a Nb over-layer $t_{eff}$ increased by 0.2 nm and for a Co over-layer by 0.5 nm. In the latter two cases there was also a substantial increase in the $O_{oxide}$/Al ratio with the larger increase occurring with the Co over-layer. There was also a shift of the peak locations to lower binding energy and a broadening of the line-widths of the oxide signals with deposition of metallic over-layers, with Co causing the larger change.

Both the over-layer and FGEB results indicate that the extent of the changes in the $AlO_x$ layer depends on the electrochemical potential gradient across the oxide. For the metal-$AlO_x$-metal samples, this is determined by the relative work functions of the two electrodes, while for FGEB it is determined by the flood-gun bias relative to the work function of the base electrode. Hence, a Co capping layer produces the larger change in the $AlO_x$ layer, while a greater flood gun bias is required to bring about a significant change on the Co-$CoO_x$ system than for Al-$AlO_x$.

The variable stoichiometry of $AlO_x$ layers, together with the low binding energy and the resultant potential instability of some of the oxygen in a completed $AlO_x$ tunnel junction, provide ready explanations for less than ideal $AlO_x$ junction performance, including drift at high bias, device-to-device variation in barrier resistance, and non-ideal junction noise behavior. On the other hand the sensitivity of the oxygen density and distribution[15] in $AlO_x$ barrier layers to the work functions of the electrodes, and to possible procedures before deposition of the top electrode, suggests pathways to improvements in tunnel barrier performance and reproducibility.

For example, 1/$f$ noise can be a major limitation to the coherence time of JJ quantum bits[16]. JJ 1/$f$ noise has been identified as arising from meta-stable atomic defect states in the barrier, whose ability to capture and release charge carriers depends on their local atomic configuration[17]. It is straightforward to identify the weakly bound oxygen ions as the origin of these unstable atomic defects. If the propensity of the as-formed thermal oxide to have a high density of oxygen vacancies arises from strain effects, any additional strain that is not due to the fundamental bond length mismatch between the oxide and the metallic electrodes should be minimized. For Nb-Al-AlO$_x$-Nb tunnel junctions, the lowest 1/$f$ noise with respect to superconducting critical current fluctuations[18] and resistance noise[19], normalized for junction and bias levels, which has apparently ever been reported was obtained when the base Nb electrode was grown so as to minimize net strain in the film. We find that FGEB of the oxide surface prior to deposition of the top electrode is another means of stabilizing the oxygen ions in the tunnel barrier. The 1/$f$ resistance noise level ($S_R/R^2$)$A$, where $\delta S_R$ is the resistance noise power spectral density, $R$ the junction resistance, and $A$ the cross-sectional area, was measured at 10 Hz at 4.2 K for 4 relatively thick Al-AlO$_x$-Al tunnel junctions (<$RA$> = 28±9 k$\Omega$·$\mu$m$^2$) that were electron bombarded before depositing the Al top electrode and compared to the noise level of a set of four junctions (<$RA$> = 13±7 k$\Omega$·$\mu$m$^2$) that were not bombarded. The former had much lower 1/$f$ resistance noise levels (<($S_R/R^2$)$A$> = (1.2±0.5)×10$^{-22}$ m$^2$/Hz) than the latter (<($S_R/R^2$)$A$> = (1.3±0.5)×10$^{-21}$ m$^2$/Hz). While the 1/$f$ noise of these Al junctions after FGEB was still much higher than the normalized noise levels of other types of well-made Josephson junctions reported in the literature[16,18,19], this reduction in 1/$f$ resistance noise level of a factor of ~10 does strongly suggest a possible pathway that might be explored for obtaining lower noise JJs and magnetic sensors that utilize $AlO_x$ tunnel barriers.

In summary, the ~300 K thermal oxidation of a thin Al layer forms an oxide layer with a nominal oxide stoichiometry of ~1.0 whose surface is covered with a layer of chemisorbed oxygen bound to the surface by positively charged oxygen vacancies. By negatively biasing the surface through a low energy electron bombardment, the chemisorbed oxygen is driven back into the oxygen vacancies. Subsequent exposure to oxygen draws these oxygen ions back onto the surface, reforming the chemisorbed layer and demonstrating the low binding energy of some of the oxygen sites in the oxide. The oxygen content and distribution in an $AlO_x$ tunnel junction is strongly affected by the work functions of the electrodes. Depositing a metallic over-layer onto the oxide will, depending on the metal work function, either result in some of the chemisorbed oxygen moving into the oxide, increasing its thickness and oxygen content, and/or in the partial oxidation of the over-layer. This offers a means of purposely tuning the oxygen in $AlO_x$ tunnel barriers through choice of electrodes or by processing of the layer before deposition of the top electrode.

This research was supported in part by the Office of Naval Research, by the Army Research Office, and by the NSF through the Cornell Center for Materials Research.

FIG. 1. XPS Al 2p and O 1s spectra of a 5 mtorr·s oxidized aluminum sample. (a) Al 2p spectra showing the metal ($Al^0$) and the oxide peaks ($Al^{3+}$), (i) as formed by a 5 mtorr·s oxygen dose, (ii) after FGEB at 20 V for 1 hour. (b) O 1s spectra, showing the oxide ($O_{oxide}$) and chemisorbed peaks ($O_{chem}$), (i) as formed by a 5 mtorr·s oxygen dose, (ii) after FGEB at 20 V for 1 hour, (iii) exposed to a 10 torr·s dose after the FGEB.

FIG. 2. Dependence of oxide film properties on oxygen exposure. Inverse of the effective oxide thickness $t_{eff}$ (top) and normalized $O_{chem}$ intensity and oxide stoichiometry (bottom) as a function of oxygen exposure for an Al thin film. The error bars represent XPS counting statistics.

FIG. 3. Effect of low energy electron bombardment. Oxide thickness and stoichiometry, initially formed by a 5 mtorr·s oxygen dose, after a each of a series of electron bombardment (using a flood gun with 20 V bias) and re-oxidation steps. The first re-oxidation (step 3) was at 300 torr·s and the second (step 5) was at $10^4$ torr·s.

FIG. 4. Effect of a capping layer on the oxide film. Oxide thicknesses of 1 torr·s oxidized Al film after capping by Y, Nb and Co over-layers. The grey band indicates the nominal thickness for an uncapped 1 torr·s oxidized film. Work function values were taken from Reference 13.

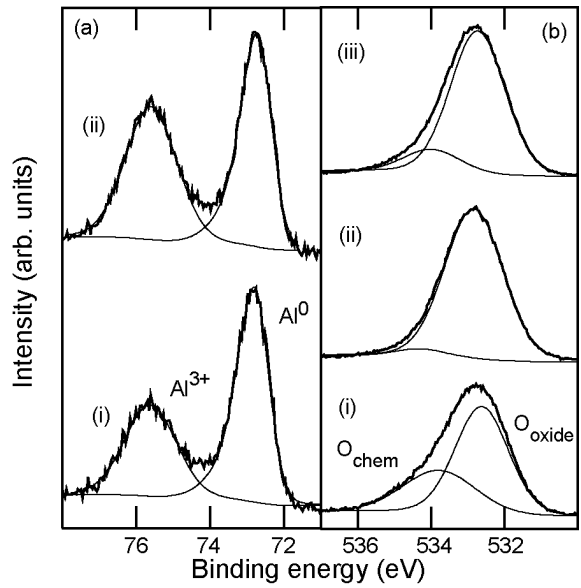

Figure 1, E. Tan

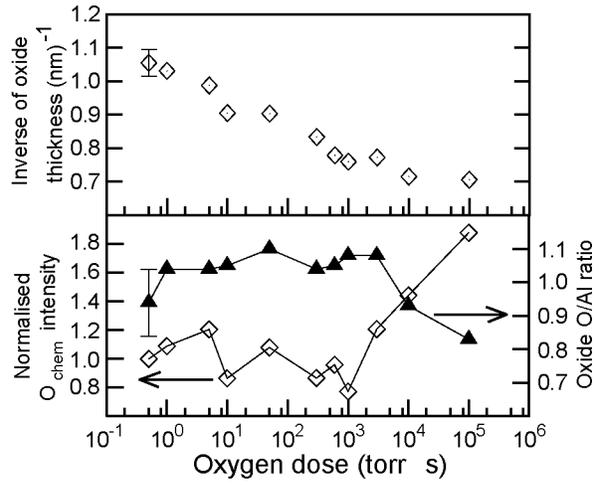

Figure 2, E. Tan

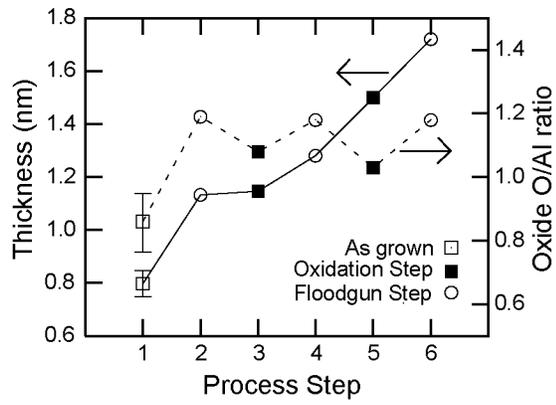

Figure 3, E. Tan

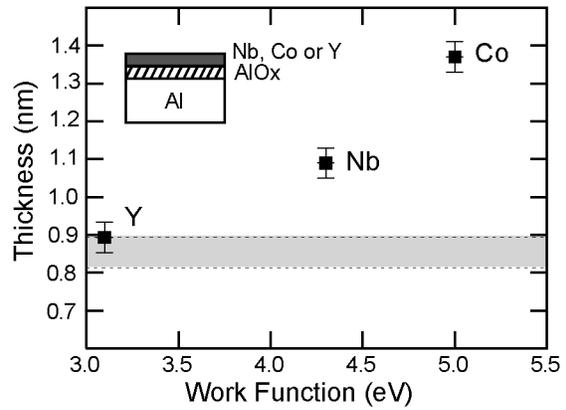

Figure 4, E. Tan

13